**Don't throw efficiency out with the bathwater: A reply to Jeffery and Verheijen (2020)**

Bartosz Bartkowski

UFZ – Helmholtz Centre for Environmental Research, Department of Economics, Permoserstraße 14, 04318 Leipzig, Germany, e-mail: [bartosz.bartkowski@ufz.de](bartosz.bartkowski@ufz.de), phone: +49 341 235 482402

**Abstract**

In this communication, I reply to the recent article by Jeffery and Verheijen (2020) 'A new soil health policy paradigm: Pay for practice not performance!'. While expressing support for their call for a more pronounced role of soil protection in agri-environmental policy, I critically discuss the two main elements of their specific proposal: its emphasis of the concept of soil health and the recommendation to use action-based payments as the main policy instrument. I argue for using soil functions as a more established concept (and thus more adequate for policy purposes), which is also informationally richer than soil health. Furthermore, I provide a more differentiated discussion of the relative advantages and disadvantages of result-based and action-based payments, while addressing the specific criticisms towards the former that Jeffery and Verheijen voice. Also, I suggest an alternative approach (a hybrid model-based scheme) that addresses the limitations of both Jeffery and Verheijen's own proposal and the valid criticisms they direct at result-based payments.

Agricultural soils provide multiple ecosystem services that positively affect human well-being (Bartkowski et al., 2018; Dominati et al., 2010; Paul et al., 2020). However, their recognition in agri-environmental policy, including in the European Union (EU), is limited (Ronchi et al., 2019). In their recent discussion article, Jeffery and Verheijen (2020) use the ongoing process of reshaping the agricultural policy of the United Kingdom (UK) post-Brexit (i.e. following UK's exit from the EU's Common Agricultural Policy, CAP) to propose a new approach for soil policy, which they call the 'new soil health paradigm'. This proposed approach has two basic tenets: first, it is based on the concept of soil health; second, its main policy instrument is supposed to be payments (akin to the agri-environment and climate measures (AECM) within the second pillar of the CAP) based on practices, not performance. While this proposal is laudable and would be a step forward against the largely soil-blind current agri-environmental policy (Bartkowski et al., 2021a), also in the EU, both key elements of the 'new soil health paradigm' (soil health as focal point and practice orientation of payments) are problematic when looked at in detail, apart from only the soil health focus being genuinely new. This reply is therefore not meant as an all-out critique of Jeffery and Verheijen's (2020) proposal – in fact, I share the basic concerns of the authors – but rather as a reformulation of their general argument. Given my disciplinary background, I will spend more time discussing the practice orientation; but also the role of the soil health concept warrants at least a brief critical reflection.

Jeffery and Verheijen emphasize the importance of soil biota for the concept of soil health as well as for the provision of soil-based ecosystem services. This emphasis is important (Soliveres et al., 2016), but at the same time points to serious limitations of the suggested 'paradigm'. While we know that soil biota are crucial for many soil processes and for the provision of soil functions (and soil-based ecosystem services), our understanding of their exact influence, the involved mechanisms and their responses to management is still quite limited (see e.g. Vogel et al., 2019; Thakur et al., 2020). From this alone, two major issues arise. First, using a soil health indicator for policy purposes is problematic because it glosses over the trade-offs between different soil processes, functions and, ultimately, soil-based ecosystem services. These trade-offs are particularly pronounced in managed soils (Schröder et al., 2020). Second, our limited understanding of the biological side of soils shows in an exemplary way why a practice-oriented (or, to use a more common term, action-based) approach, at least as proposed by Jeffery and Verheijen, is problematic – it would be naïve to assume that we can identify a set of practices that support soil health (i) everywhere (irrespective of spatial heterogeneity of both soils and their management) and (ii) in all respects (individual soil functions).

This is aggravated by the fact that, as of now, there is no agreed-upon soil health indicator available (Lehmann et al., 2020), as Jeffery and Verheijen admit themselves. This does not by itself put into question the epistemic value of the soil health concept, but for practical and especially policy purposes, more established concepts are required. In fact, in the context of soils, we have such a concept at our disposal – soil functions (Baveye et al., 2016; Vogel et al., 2018), which furthermore have the additional benefit of being clearly relatable to human well-being via the concept of ecosystem services (Adhikari and Hartemink, 2016; Paul et al., 2020). Also, by jointly looking at multiple individual indicators, rather than at one composite soil health indicator, the trade-offs involved in soil management can be taken into account in an explicit and transparent way.

So, what kind of instruments can and should be used to advance soil health (either, as suggested by Jeffery and Verheijen, as a 'composite' or, as suggested here, via the intermediate concept of soil functions)? Jeffery and Verheijen seem to claim that the current paradigm in agri-environmental policy is performance-based (or, again to use a more common term, result-based). This claim may be true for the current rhetoric at the EU level (and beyond), where the importance of paying for results, not actions, is emphasized – and for good reason, e.g. to take into account heterogeneity and information asymmetries, to provide incentives for innovation and to increase cost-effectiveness of payment schemes (see Burton and Schwarz, 2013; Cullen et al., 2018; Hasund, 2013; White and Hanley, 2016). However, when looking at the spectrum of AECM implemented within the CAP across member states, very few are result-based (Herzon et al., 2018).[1] This is related to a challenge correctly pointed out by Jeffery and Verheijen – result-based payments require extensive monitoring and measurement of the 'results', which is challenging for most environmental public goods, including soil functions.[2]

However, this does not mean that conventional action-based payments are the solution. One of the major challenges for action-based payments, which is particularly pronounced in the context of soils, is spatial heterogeneity. This heterogeneity may be related to both environmental conditions and management and behaviour of land-users. As a result, different practices will mostly be effective in different places and in different management contexts, and also the trade-offs between soil functions are likely to vary spatially (Schröder et al., 2020). Therefore, it is necessary to provide incentives for spatial allocation of measures where they are (most) cost-effective – something that result-based payments do quite well. Another possible alternative approach is to use advances in (joint) modelling of soil functions, where the model(s) can provide spatially highly resolved information about the changes in soil functions under different management scenarios (Vogel et al., 2018). Farmers could then be remunerated on the basis of these modelled 'results' specific to their plots and their particular management context (Bartkowski et al., 2021b).

Are such 'payments by modelled results' action-based or result-based? This probably is not a sensible or practically relevant question. It should be acknowledged that the main limitation of model-based payments as compared to conventional result-based payments is that the latter provide strong dynamic incentives for innovation by farmers, which a model-based scheme can only do in an indirect and imperfect way (Bartkowski et al., 2021b). It is worth emphasizing that, apart from measurement challenges, result-based payments should be the approach to be strived for and approximated, while action-based are nothing more than a pragmatic compromise. The reasons for this have been elaborated in detail elsewhere (e.g. Bartkowski et al., 2021b; Burton and Schwarz, 2013). Here, I would like to focus on specific criticisms that Jeffery and Verheijen direct at the result-based approach and provide a rebuttal of some of the arguments they make. Specifically, they make two claims about supposed limitations of result-based payments that, while not completely false, deserve a more differentiated discussion.

---

[1] A comprehensive, continuously updated collection of information about result-based payment schemes across Europe is provided by the Result Based Payments Network, https://www.rbpnetwork.eu/
[2] Nonetheless, there exists at least one pilot result-based payment scheme targeting soils – the "Klimaschutz durch Humusaufbau" programme in the Swiss canton Basel-Landschaft, which started in 2019 and where payments are based on measured changes in soil organic matter.

The first limitation is what Jeffery and Verheijen call 'slay for pay', i.e. the risk that rational farmers will deliberately downgrade their fields in order to achieve significant increases in soil quality (measured in terms of soil health, soil functions or else) easier. This possible behavioural pattern is an instance of adverse selection (in economic nomenclature), i.e. the inability of the regulator to distinguish between good and bad behaviour before the payment contract is signed. A related problem is that farmers may adopt sustainable practices on enrolled plots, but compensate this by intensifying management on other plots (leakage). These are real dangers, though one must also keep in mind that even on a 'slain' plot, soil quality improvements are not cost-free, so the incentive for farmers to engage in such problematic behaviour may not be as strong as suggested by Jeffery and Verheijen. Ultimately, this is an empirical question. However, the danger of 'slay for pay' should be evaluated not in isolation, but in comparison to the alternative, i.e. action-based payments. Here, another form of adverse selection is present in conventional, not spatially targeted action-based schemes, where rational farmers (and Jeffery and Verheijen assume such themselves when they invoke the danger of 'slay for pay' behaviour) will select plots for the scheme not based on where the soil function/soil health effects will be highest, but where it will cost them least (in terms of income foregone, effort etc., i.e. opportunity and transaction costs). This may heavily impair the effectiveness of an action-based scheme. So, instead of throwing result-based schemes out with the bathwater, it may be more useful to consider how to address the 'slay for pay' danger in the design of the scheme. One option could consist in using some general baseline, such as 'good agricultural practice', instead of the actual status quo condition of soil at the plot to be enrolled (Bartkowski et al., 2021b). The downside would be the potential inefficiency due to paying for non-additional effects (windfall gain) if the actual status quo is above the hypothetical baseline. Especially in a model-based scheme design, such hypothetical baseline based on 'typical' management (but given the site-specific natural conditions of the plot[3]) may actually be more easily implemented than the alternative of measuring the status quo.

The other criticism of result-based payment schemes voiced by Jeffery and Verheijen concerns fairness considerations. They argue that "a payment system based on measured soil health improvements is inherently biased against farms with already healthy soils" (Jeffery and Verheijen, 2020, p. 372). In other words, unincentivized early adoption of beneficial practice is 'penalized' in a result-based scheme (note that this is largely independent of the baseline choice discussed above). However, this argument overlooks that unincentivized early adoption can have different reasons, ranging from intrinsic motivation to relatively favourable natural conditions or relatively large action space of a farm, e.g. in terms of capital availability. In light of this, whether not paying those early adopters is 'unfair' very strongly depends on the particular context and the reasons behind their early adoption. Furthermore, it equally strongly depends on the applied fairness criterion, of which there are many, sometimes contradicting ones (Elster, 1992). Given heterogeneous mitigation costs (both implementation and opportunity costs of soil benefitting practices), it may actually be fair to remunerate only the relatively disadvantaged farmers who were not in the position for early adoption. In the end, as perceptions of fairness are highly context-specific and vary across communities, they should be considered in the scheme design via involvement of stakeholders.

Furthermore, even if we grant the unfairness argument, there seems to be a trade-off between 'fairness' and efficiency, as paying early adopters violates the criterion of additionality and thus leads to waste of scarce public resources (Engel, 2016). This trade-off, however, is based on a simplistic, overly static interpretation of additionality. Actually, additionality is a criterion that compares the actual state of the world where the payment is provided with a counterfactual, in which it is not. Even though it may be non-additional and thus inefficient to remunerate an early adopter for something she already did without the remuneration, there is no guarantee she won't reverse management in the future in response to adverse developments with respect to markets, climate, peer pressure or other factors. Therefore, an additionality-compatible argument could be made for a double-tier scheme with payments for adoption (tier 1) and payments for maintenance (tier 2), where the latter would probably be lower to reflect that initial investment had already been undertaken and that the counterfactual with management reversal is uncertain. Such scheme could be a variation of a base payment/result-based payment hybrid scheme as discussed by Derissen and Quaas (2013), though in our case, the base payment need not be action-based. Some form of incentives for maintenance of beneficial practices would generally be sensible given the long-time horizons involved in sustainable soil management.

---

[3] I.e. the hypothetical baseline would be based on site-specific inherent soil properties combined with 'typical management parameters', instead of measuring the actual realization of the manageable properties of the site (Vogel et al., 2019).

To sum up, Jeffery and Verheijen are right in pointing out the need to more strongly consider soils in agri-environmental policy. It is doubtful, however, that the advocated soil health concept provides the most sensible framework to do that. Furthermore, while Jeffery and Verheijen's criticism of the result-based payment approach raises important design questions, the advocated shift to (or, actually, sticking to) action-based payments would likely constitute a proverbial act of throwing the baby out with the bathwater.

**Acknowledgements**

I would like to thank Hans-Jörg Vogel for helpful discussions. The usual disclaimer applies. This work was funded by the German Federal Ministry of Education and Research (BMBF) in the framework of the funding measure "Soil as a Sustainable Resource for the Bioeconomy – BonaRes", project "BonaRes (Module B): BonaRes Centre for Soil Research, subproject A" (grant 031B0511A).